# Laser-induced deposition of single-walled carbon nanotubes in suspended core optical fibers

Ricardo E. da Silva[1*], Hartmut Bartelt[2], and Cristiano M. B. Cordeiro[1]
[1]*Institute of Physics Gleb Wataghin, University of Campinas*, Campinas, Brazil
[2]*Leibniz Institute of Photonic Technology (IPHT), Jena, Germany*
*resilva@unicamp.br

*Abstract*— We experimentally demonstrate laser-driven deposition of single-walled carbon nanotubes (CNTs) in suspended core fibers (SCFs) for the first time. Two samples for each SCF type with three (SCF1) and four (SCF2) air holes are individually inserted in a syringe loaded with a 0.5 mL solution of CNTs dispersed in methanol, and a high-power laser at 980 nm is injected into the fibers during distinct periods: SCF1 (20 and 30 minutes), SCF2 (5 and 10 minutes). The CNT solution thermally expands, accelerating and depositing nanotubes on the fiber tip. The material and dimensions of the CNTs deposited on the SCFs' cross-section are characterized, revealing nanotube bundles increasing in thickness from the fiber core to the bridges and inside the holes. The average maximum CNT thickness widening the bridge edges is, respectively, for SCF1 (2.8 and 2 μm) and SCF2 (1.7 and 1.3 μm). These achievements are significant for nanoparticle deposition and fabrication of microscale devices on tips and inside specialty and hollow fibers, enabling innovative fiber sensors, mode-locked fiber lasers, in-fiber acousto-optic modulators, and broadband photoacoustic transmitters for imaging and neurostimulation.

*Keywords—Laser-induced deposition, carbon nanotubes, suspended core fibers, in-fiber CNT-silica devices.*

## I. INTRODUCTION

All-fiber devices integrating carbon nanotubes (CNT) enable interesting applications in sensors, photonic switches for communications, industrial mode-locked pulsed fiber lasers, and photoacoustic ultrasonic devices for imaging and neurostimulation in biomedicine [1], [2], [3], [4]. CNTs provide remarkable chemical, mechanical and optical properties, including a broadband absorption spectrum for modulating high-power femto-to-picosecond fiber lasers [5], [6]. However, increasing thickness of CNT composites deposited in the fiber core might increase non-saturable losses, reducing the laser power damage threshold and power output [6]. In contrast, depositing CNTs as rings around the fiber core, tapered fibers, D-shaped fibers or inside photonic crystal and hollow core fibers are promising to overlap propagating modal evanescent fields and CNTs, reducing losses and damage in the fiber core while improving device lifetime and laser power output [1], [7], [8]. Carbon deposition on fiber tips can generate highly confined ultrasound by thermal expansion of CNT composites and is used, e.g., for neurostimulation investigating brain diseases [4]. In this case, thick nanotube layers generate high-pressure ultrasound, while thin layers increase frequency bandwidth and spatial resolution [9]. Thus, controlling the thickness, shape, and distribution of CNTs impacts device operation, efficiency, and performance.

Several techniques have been successfully proposed to deposit CNTs in optical fibers, including wallpaper vertical solvent evaporation [6] and dip coating the fiber with CNTs functionalized with organic gels [9]. For specialty and hollow fibers, an aqueous CNT solution is inserted in the air holes under gas pressure, further keeping the fibers at over room temperatures (30 °C) for a few days to evaporate the fluid, resulting in dried CNT films deposited on the air hole walls. Alternatively, laser-induced radiation offers fast and controlled CNT distribution over the fiber cross-section [1], [2], [3], [5]. The laser is injected into a fiber dipped in a CNT solution, attracting and depositing the nanotubes on the fiber tip. CNT bundles can be deposited in the fiber core, around the core, or over the entire fiber cross-section by adjusting the laser power, wavelength, duration and beam profile, and CNT-solvent properties [1], [3]. However, CNTs in the fluid moving towards the fiber are usually restricted inside the laser beam cone with power axially decreasing in the fluid from the fiber core [7]. Also, nanoparticle deposition is usually limited by competitive physical mechanisms: convection and thermophoresis expanding the fluid and moving CNTs to the laser axis, while strong radiation pressure in the core deviates the nanotubes to the cladding, edges and outside the fiber [2], [3], [7].

Here, we experimentally deposit single-walled CNTs in suspended core fibers (SCFs) composed of three and four air holes using a syringe-based laser technique for the first time. The material and geometric properties of CNT layers deposited over the SCFs' cross-section are characterized with unprecedented spatial resolution. We show that increasing the interaction between nanoparticles, the laser beam, and the SCFs inside the syringe cavity, the pressures induced by accelerating particles overcome radiation pressures from the fiber core, allowing CNT bundles to attach in the SCF bridges and air holes. The fabrication of a CNT horn-like structure and a CNT coated concave structure in the four-hole SCF is discussed, highlighting a promising technique for the fast and controllable deposition and fabrication of CNT structures in specialty optical fibers.

## II. EXPERIMENTAL SETUP AND METHODS

Fig. 1(a) illustrates the experimental setup used to deposit CNTs in the SCFs. A liquid solution with about 2.5 mg/mL concentration of single-walled carbon nanotubes (0.84 nm average diameter and carbon purity ≥ 95 %) dispersed in methanol (purity ≥ 99.8 %) is prepared using the methods described in [10]. A 3 mL maximum capacity glass syringe is loaded with a 0.5 mL solution volume and fixed on an XYZ micrometer positioner for each SCF sample.

Fig. 1(b) shows details of the inner cross-section of the employed SCFs with three (SCF1) and four (SCF2) air holes, indicating the respective average dimensions of the fiber core (6.4 and 5.1 μm) and hole diameters (57 and 24 μm), and bridge thickness (2.6 and 0.78 μm). SCF1 and 2 outer diameters are, respectively, 127 and 100 μm. Additional details about the SCFs' optical modes, confinement, and propagation are described in [11], [12]. The SCF tip is cleaved and further cleaned by applying low-power arc voltage using a fusion splicer. The SCF is then fixed on another XYZ positioner and centrally aligned to the syringe's aperture with the assistance of a digital microscope connected to a laptop. This is shown in Fig. 1(c) before inserting the SCF sample in the loaded syringe tip. Fig. 1(d) shows the SCF aligned inside the syringe before loading the CNT solution (the fiber is no longer seen when the syringe is loaded).

The other SCF end is spliced to a single-mode fiber (SMF) connected to a coupler. The laser centered at 980 nm is split, delivering distinct maximum powers at each SCF type output (~ 61 mW for SCF1) and (110 mW for SCF2) due to distinct splice losses (and SCF-SMF mode field mismatch). The power meter monitors the power at the SCF output by measuring a low-power signal from the other coupler's arm. CNT deposition is investigated by applying longer laser periods for SCF1 (20min) and SCF1 (30min), and shorter periods for SCF2 (5min) and SCF2 (10min). The longer laser periods for SCF1 aimed to compensate for its lower power output. The other setup parameters are kept the same for the samples. The setup is covered with a removable shelter, blocking laser emissions through the syringe and isolating the components and the CNT solution from contamination.

The laser injected into the syringe is partially absorbed by the CNTs which vibrate and thermally expand the surrounding fluid, generating a particle flow in the syringe (yellow curves in the container in Fig. 1(a)). The flow further converges to the fiber axis accelerating nanotubes to the SCF tip with increasing pressure magnitude for longer laser duration [14].

An electron scanning microscope (SEM) is used to image the CNT bundles deposited over the SCF's cross-section, providing microscale details of layers in the fiber core and bridges. Although it is not shown, we characterized the material properties of the CNT structures formed in different positions at the SCF's cross-section, by using an energy dispersive spectrometer (EDS - Oxford X-Max 50) connected to SEM (EDS is used to identify the material chemical elements by detecting X-rays emitted from localized regions in the deposited layers). The resulting compound materials are indicated with arrows in the images in the following section.

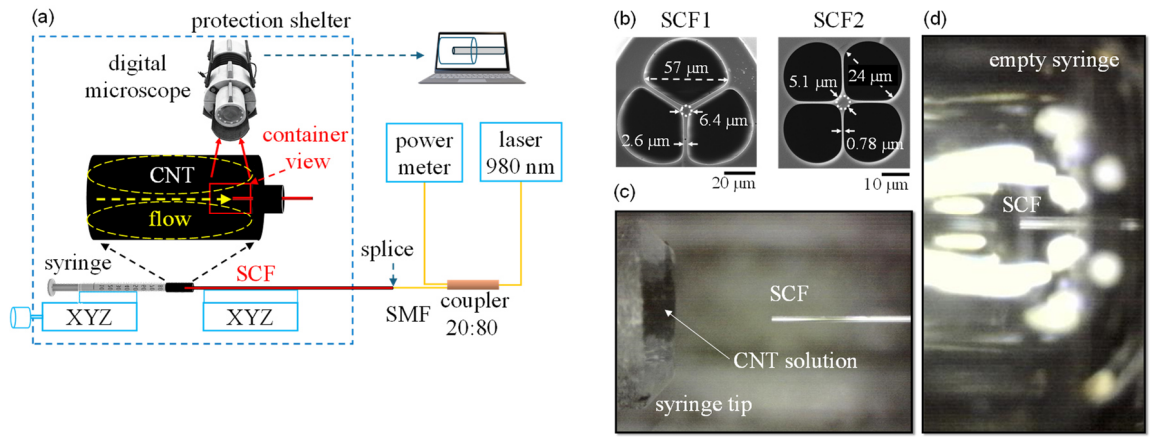

Fig. 1. (a) Illustration of the experimental setup used to deposit carbon nanotubes (CNT) on the tip of suspended core fibers (SCFs) composed of (b) three (SCF1) and four (SCF2) air holes. (c) Syringe tip loaded with the CNT-methanol solution before inserting an arbitrary SCF. (d) An SCF aligned inside the syringe before loading the CNT solution.

## III. RESULTS AND DISCUSSION

Fig. 2(a) shows SCF1 (20min) with CNTs distributed over the fiber cross-section (the inset shows a core detail of a clean SCF for comparison). These layers are mostly composed of carbon with negligible impurities (mostly salt-shaped minerals probably from CNT powder, solvent, SCF, and employed recipients, and should not affect optical and ultrasonic applications [7], [9]). Fig. 2(b) shows the fiber core with CNT bundles highly concentrated in the surrounding bridges (the average maximum thickness of CNTs widening the bridges is 2.8 μm). This indicates that the nanoparticles are partially deposited and partially scattered to the bridges, which attract and accumulate bundles in the optical evanescent fields around the core [13] (Fig. 2(c) shows the left bridge with a dense CNT agglomeration in the inset). Fig. 2(d) shows a detail in the fiber core, indicating a thin layer of CNT bundles compared to those deposited on the bridges.

Fig. 2(e) shows SCF1 (30min). We note that CNTs are mostly concentrated over the fiber core and bridges, indicating that longer laser exposure engages more particles in the flow, accelerating nanotubes towards the fiber core. Consequently, CNTs colliding with the core overcome radiation pressures, focusing nanotubes mostly on the core and bordering bridges compared to SCF1 (20min) (the average maximum thickness of CNTs coating the bridge edges is 2 μm in Fig. 2(f)).

Fig. 2(g) shows a detail of the right-bridge with a high concentration of nanotubes (inset). A careful observation in Fig. 2(h) reveals nanotube bundles impressively growing in diameter over the core surface. Overall, CNT bundles are also observed coating internally the bridges in Fig. 2(c) and 2(g), indicating layers deposited along the fiber length for both SCF samples (deep layers could not be characterized with our current resources). Overall, CNTs deposited along the fiber should increase the evanescent saturable optical absorption as those offered in D-shaped fibers, fiber tapers, and hollow core fibers [7].

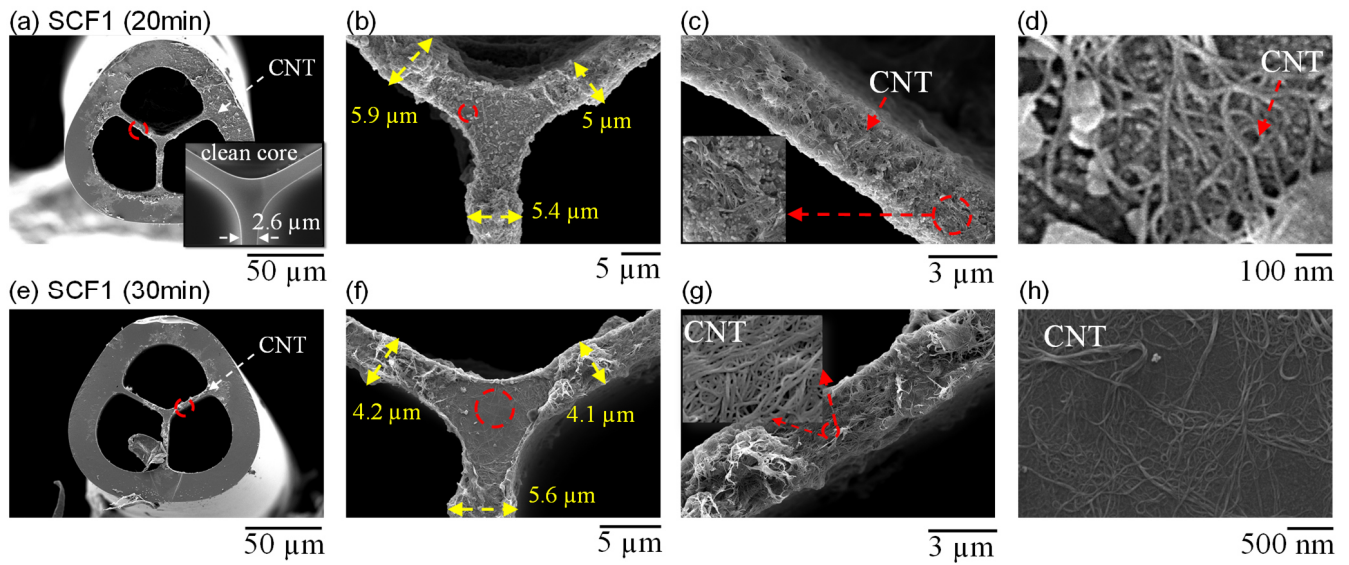

**Fig. 2**. (a) SCF1 (20min) and (e) SCF1 (30min) cross-sections with details of carbon nanotube (CNT) bundles, respectively (b)(f) in the core and selected (c)(g) silica bridges indicating dense nanotube agglomeration (insets) in the cladding and (d)(h) inner regions in the fiber core in both SCF samples.

Fig. 3(a) shows SCF2 (5min) with CNTs deposited over the fiber cross-section. The carbon bundles increase from the fiber core to the bridges, strongly accumulating in the right air holes, as shown in the upper hole in Fig. 3(b). Fig. 3(c) shows particles mostly composed of silica and carbon deposited in the core, denoting superficial ablation in the core region. This should be caused by the increasing formation of shock waves striking with high velocities and pressures towards the fiber, removing and spreading silica around the core [14].

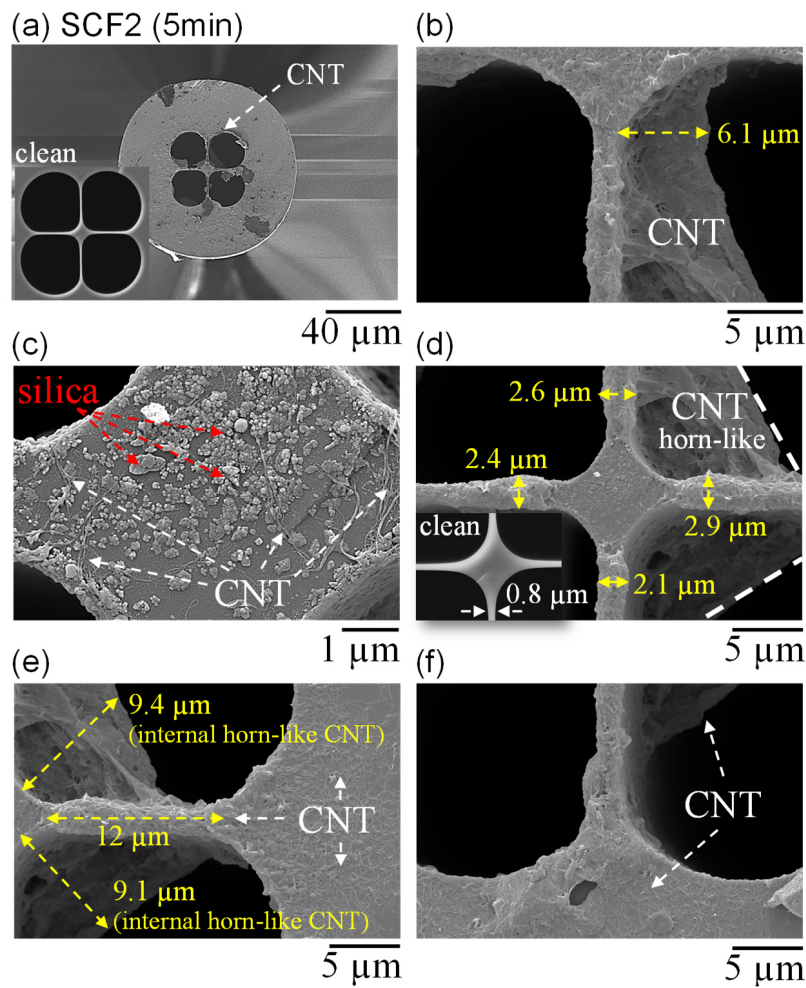

**Fig. 3**. (a) SCF2 (5min) cross-section with details of carbon nanotube (CNT) bundles increasing from the fiber core to the (b) bridges and accumulating in the right holes. Detail of (c) the fiber core and (d) CNT bundles in the bridges forming a (e) internal horn-like carbon structure. (f) CNTs expand to the outer cladding around the holes. Insets in (a) and (d) show clean SCF cross-sections.

In contrast, the evanescent fields along the bridges strongly attract the nanotubes, accumulating bundles in the holes expanding to the outer cladding in Fig. 3(d)-(f) (average maximum CNT thickness on the bridge edges is 1.7 μm in Fig. 3(d)). The nanotubes in the holes form an internal triangular horn-like carbon structure (outer edge indicated with white dashed line in Fig. 3(d)), expanding about 9 μm from the core border, as shown in Fig. 3(e). The deposition non-uniformities (and strong CNT deposition mostly in the right holes) might be caused by the nonsymmetric distribution of higher-order modes propagating at the laser wavelength (980 nm) [11], [13]. Overall, this horn-like structure is promising for the fabrication of in-fiber acousto-optic modulators, potentially generating highly efficient and high-frequency confined ultrasonic waves inside SCFs to modulate active fiber lasers [11].

Fig. 4(a) shows SCF2 (10min) with the concave structure formed in the inner cladding (the red circle in Fig. 4(b) indicates the depressed region of the fiber core and bridges compared to the outer silica cladding). In addition, CNTs widen the bridges (average 1.3 μm) and expand to the hole borders, as shown in Fig. 4(c) and 4(d). This concave profile might be caused by highly concentrated nanoparticles colliding with the fiber core and bridges for a longer laser duration (Fig. 4(e) shows particles mostly composed of silica, indicating superficial ablation in the core). Overall, the resulting CNT coated concave structure is apparently uniform, which is promising to generate and focus ultrasonic pulses to reach micro-targets in photoacoustic imaging and neurostimulation [4], [9].

In summary, SCF1 and 2 structures show a nearly uniform distribution of nanotube bundles over the fiber core and along the bridges. Further studies should consider lasers approaching 1550 nm to reduce multimode-induced asymmetries and deposition in the bridges, while enhancing fundamental mode and deposition in the core [12] (deposition non-uniformities might also come from fiber misalignments and tilts, or instabilities in the fluid flow [14]). Overall, controlling the propagating modes in the SCFs should potentially tune the overlap of optical power and deposited CNTs, regulating or switching the power absorption between the fiber core and bridges with distinct carbon thickness. These features should reduce absorber losses and power damage in the core, improving power output and efficiency of mode-locked fiber lasers.

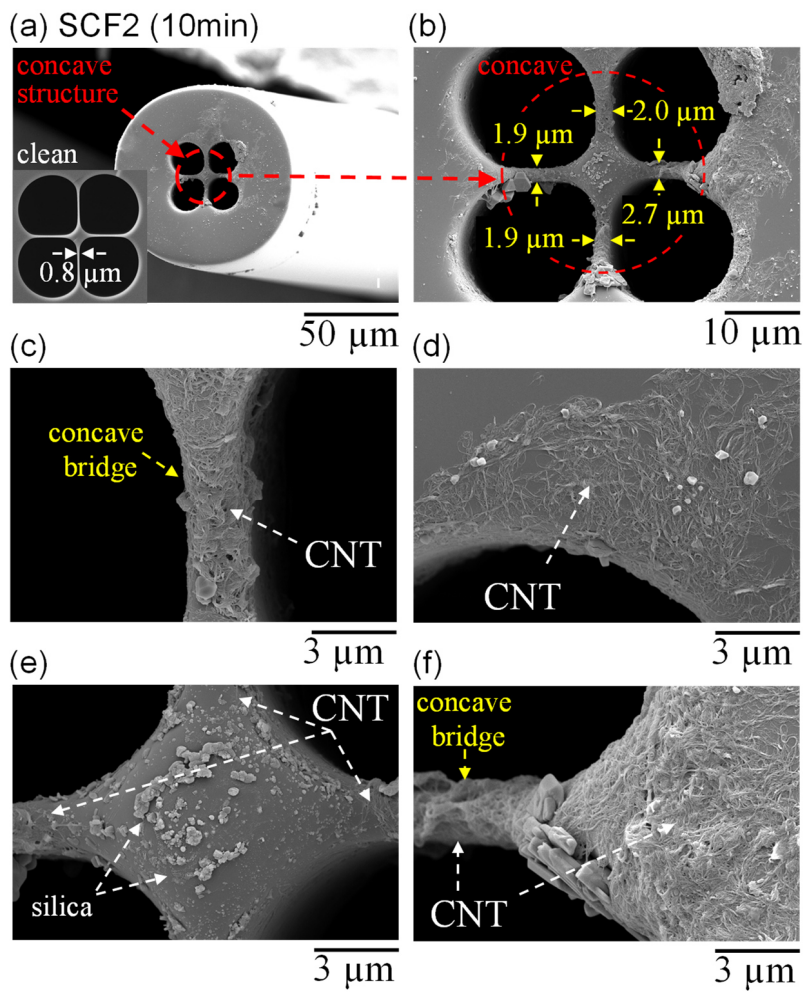

**Fig. 4**. (a) SCF2 (10min) cross-section with details of the (b) concave structure with carbon nanotube (CNT) bundles increasing concentration from the (c)-(f) fiber core to the bridges and outer silica cladding (inset: clean cross-section).

## IV. CONCLUSION

We have experimentally demonstrated the deposition and fabrication of CNT structures in suspended core optical fibers for the first time. SCFs with three and four air holes are inserted into a CNT-methanol solution and excited, respectively, with laser powers of 61 and 110 mW over distinct durations SCF1 (20 and 30 minutes), and SCF2 (5 and 10 minutes). In general, CNT bundles increase from the fiber core to the bridges, which is promising for power transmission while absorbing and modulating evanescent optical fields. Thus, shorter laser periods increase nanotube distribution in the fiber core and overall fiber cross-section, while longer periods focus high-pressure particles depositing thinner layers in the fiber core. SCF2 (5min) shows a CNT horn-like structure inside the holes, while SCF2 (10min) reveals a concave structure coated with nanotubes. These achievements are promising for microscale high-frequency in-fiber active acousto-optic modulators, saturable absorbers for passive mode-locked fiber lasers, and the generation of broadband, highly focused ultrasound pulses in biomedical applications.

## ACKNOWLEDGMENT

This work was supported in part by the grants 2022/10584-9, 2024/00998-6, São Paulo Research Foundation (FAPESP), 309989/2021-3, Conselho Nacional de Desenvolvimento Científico e Tecnológico (CNPq). We thank Prof. João P. V. Damasceno with the Group of Nano Solids (GNS) for kindly providing the CNT solution. We thank the groups with the LAMULT – IFGW (Rosane Palissari, Bruno Camarero), and LIMicro – IQ (Hugo Campos Loureiro and Ana Letícia Moreira da Fonseca) for the support to characterize the SCF samples (Microscopy Core Facility RRID:SCR_024633 at UNICAMP for the support - the Quanta FEG 250 system was partially funded by a FAPESP grant #2023/01620-4).